\documentstyle[aps,twocolumn]{revtex}
\begin{document}
\title{Barrier resonances in Bose--Einstein condensation}
\author{Juan J. G. Ripoll and V\'{\i}ctor M. P\'erez-Garc\'{\i}a}

\address{Departamento de Matem\'aticas, Escuela T\'ecnica Superior de
  Ingenieros Industriales\\ Universidad de Castilla-La Mancha, 13071
  Ciudad Real, Spain}

\maketitle

\begin{abstract}
  We study the dynamics of the mean field model of a Bose--Einstein
  condensed atom cloud in a parametrically forced trap by using
  analytical and numerical techniques. The dynamics is related to a
  classical Mathieu oscillator in a singular potential.  It is found
  that there are wide resonances which can strongly affect the dynamics
  even when dissipation is present. Different geometries of the forcing
  are discussed as well as the implications of our results.
\end{abstract}


\date{\today}

\narrowtext

\section{Introduction}

The recent experimental realization of Bose--Einstein condensation (BEC)
in ultra-cold atomic gases, \cite{Science,Hulet} has triggered the
theoretical exploration of the properties of Bose gases. Specifically
there has been a great interest in the development of applications which
make use of the properties of this new state of matter. Perhaps, the
recent development of the so--called atom laser \cite{atomlaser} is the
best example of the interest of these applications.

The current model used to describe a system with a fixed mean number $N$
of weakly interacting bosons, trapped in a parabolic potential
$V(\vec{r})$ is the following Nonlinear Schr\"odinger Equation (NLSE)
(which in this context is called the Gross--Pitaevskii equation (GPE))
\begin{equation}
  \label{pura}
  i \hbar \frac{\partial \psi}{\partial t} =
  -\frac{\hbar^2}{2 m} \nabla ^{2} \psi +
  V(\vec{r},t)\psi + U_0 |\psi|^2 \psi,
\end{equation}
which is valid when the particle density and temperature of the
condensate are small enough. Here $U_0 = 4 \pi \hbar^2 a/m$
characterizes the interaction and is defined in terms of the ground
state scattering length $a$.  The normalization for $\psi$ is $N = \int
|\psi|^2 \ d^3 \vec{r},$ and the trapping potential is given by
\begin{equation}
  \label{parabolic}
  V(\vec{r},t) = \frac{1}{2} m\nu^2
  \left( \lambda_x^2(t) x^2 + \lambda_y^2(t) y^2
    + \lambda_z^2(t) z^2 \right),
\end{equation}

The $\lambda_\eta, \ (\eta = x,y,z)$ are, as usual, functions that
describe the anisotropies of the trap \cite{Dalfovo}. In real
experiments with stationary systems they are constants and the geometry
of the trap imposes usually the condition $\lambda_x= \lambda_y=1$.
$\lambda_z=\nu_z/\nu$ is the quotient between the frequency along the
$z$-direction $\nu_z$ and the radial one $\nu_r \equiv \nu$.

At this point we want to emphasize here that our analysis does not
restrict to the stationary case. Instead it focuses on situations where
$\lambda_\eta$ are periodic sinusoidal functions of the time. To be more
precise we will adopt the notation
\begin{equation}
  \label{lambdas}
  \lambda_i(t) = \lambda_{i,0}(1 + \epsilon_i cos(\omega_i t))
\end{equation}
with $i = x,y,z$.

In its original derivation Eq. (\ref{pura}) was supposed to be strictly
valid in the $T=0$ and low density limits. Recent theoretical work
extends the applicability of the GPE to the high density limit
\cite{Gard,Ziegler}. On the other hand linearized stability analysis
based on perturbative expansions on $1/N$ seems to point that the
validity of the equation is restricted to the cases where no exponential
separation of nearby orbits appear as it happens for example in chaotic
pulsations of the atom cloud \cite{Castin2,Castin3}.

A lot of work has concentrated in the analysis of the resonance
structure when a periodic time dependent perturbation is applied to the
magnetic field, mainly because of the availability of experimental data
\cite{expfreq}. The theoretical advances include both semi-classical
analysis of Eq. (\ref{pura}) in the hydrodynamic limit \cite{Stringari},
variational methods \cite{theorfre,Victor2} and numerical simulations
\cite{Ruprecht1,Ruprecht2}.  Recent research based on Eq. (\ref{pura})
also includes the study of the different problems using tools from
nonlinear science \cite{Bolitons,Tsurumi1,otros} and the analysis of its
multicomponent extensions \cite{multiple1,multiple2}.

The variational approach provides us with very simple equations for the
evolution of the widths of the Gaussian atom cloud together with a
simple picture of the movement of the condensate \cite{Victor2}.
However, those equations were initially derived for the static field
case and only applied to the analysis of the normal modes and
frequencies of the condensate motion in the weak perturbation regime.
Similar equations are found using various scaling arguments
\cite{Kagan,Castin1,Castin2} or moment theory \cite{multiple1}.

Evolution of the condensate in a time dependent trap has been addressed
in many different ways.  There are papers where the GPE is solved
numerically, either by watching the time evolution of a suitable initial
condition \cite{Ruprecht1,Ruprecht2} or by using a linear expansion in
normal modes \cite{Ruprecht2}. And there are other works
\cite{Kagan,Castin2} where the authors derive a set of ordinary
differential equations using scaling arguments plus the Thomas--Fermi
approximation.  Among all the papers that treat time dependent
potentials, the one which is closer to what we present here is
\cite{Castin2}. In that work the authors study the excitations of
condensed and non--condensed clouds under a periodic parametric
perturbation and find that the condensate depletes and that chaotic
evolution comes out. We will comment more on this at Sec. \ref{Conclu}.

It is our intention in this paper to explore the behavior of the
condensate under periodic perturbations of the trap strength using
analytical and numerical tools.  Analytical techniques include exact
results and a variational analysis extended to situations with a time
dependent potentials. Comparison of the results of both methods will
allow us to state rigorously and derive a simple model for the resonant
behavior and to qualitatively predict the evolution of the system under
other conditions (less symmetry, dissipation regimes, non--condensed
corrections, etc).  Other conclusions as well as experimental
implications will be found in the course of the analysis.

Our plan is as follows: In Sec. \ref{Center} we obtain a set of exact
equations for the evolution of the center of mass of the
condensate. The result is found to be three uncoupled Mathieu
equations.  We show that a complete perturbative analysis is possible
and obtain the resonance structure for the center of mass.  In Section
\ref{Variational} we introduce the time dependent variational model
for the time dependent GP equation.  In Sec. \ref{Radial} we solve the
GPE numerically in the radially symmetric case. We show the resonance
structure for the condensate in all regimes -- small and large
amplitude oscillations --. Next we turn to the variational equations
for the widths and show that numerical simulations agree qualitatively
with the preceding picture. Finally, we use those ordinary
differential equations to predict the locations of the resonances in a
simple model that connects the linear and the nonlinear cases using
tools from Sec.  \ref{Center}. This is the main result of our
paper. In Sec. \ref{Nonsym} we consider the possibility of extending
the analysis to the multidimensional case and study the effect of the
coupling between the different variational parameters. The resulting
predictions are found to agree with a posterior set of 3D simulations
of the GPE equation. In Sec. \ref{losses} we comment on the effect of
losses. We show that these resonances are of a persistent nature and
show how chaotic evolution may arise from losses effects. Finally in
Sec. \ref{Conclu} we discuss the experimental implications of our
results and summarize our conclusions.

Note that, unless otherwise stated, all magnitudes are in adimensional
units. To adimensionalize these figures we have employed the change of
variables that is introduced in Sec. \ref{Variational}

\section{Exact analytical results}
\label{Center}

\subsection{Variational form of the GP equation}

It can be proved that every solution of Eq. (\ref{pura}) is a
stationary point of an action corresponding, up to a divergence, to the
Lagrangian density
\begin{eqnarray}
  \label{density} {\cal L} & = & \frac{i\hbar}{2} \left(
  \psi\frac{\partial \psi^{\ast}}{\partial t} - \psi^{\ast}
  \frac{\partial \psi}{\partial t} \right) \nonumber \\ & + &
  \frac{\hbar^2}{2m} |\nabla \psi|^2 + V(r) |\psi|^2 + U_0 |\psi|^4,
\end{eqnarray}
where the asterisk denotes complex conjugation. That is, instead of
working with the NLSE we can treat the action,
\begin{equation}
\label{action}
  S = \int {\cal L} d^3rdt = \int_{t_i}^{t_f} L(t) dt,
\end{equation}
and study its invariance properties and extrema, which are in turn
solutions of Eq. (\ref{pura}).

For instance, from the invariance of Eq. (\ref{density}) under global
phase transformations, one can assure the conservation of the norm of
the wave function
\begin{equation}
  \label{norm}
  N = \int |\psi|^2 d^3r,
\end{equation}
which in this context is interpreted as the number of particles in the
Bose condensed state. We can also define another quantity,
\begin{equation}
  \label{energy}
  E = \int \left\{\frac{\hbar^2}{2m}|\nabla\psi|^2 + V(r,t)\psi
    + \frac{U_0 |\psi|^4}{2} \right\} d^3r,
\end{equation}
that can be thought of as the energy, and whose time evolution is
simply
\begin{equation}
  \label{energy-dot}
  \frac{dE}{dt} = \int \frac{dV}{dt} |\psi|^2 d^3r.
\end{equation}
Thus, when the potential is not time dependent, the energy is another
conserved quantity. And when the potential has the form
(\ref{parabolic}), the evolution of energy can be easily connected to
that of the mean square radii of the cloud
\begin{equation}
  \label{energy-dot-here}
  \frac{dE}{dt} = \frac{1}{2}m\nu^2
  \sum_{\eta = x,y,z} \frac{d\lambda_\eta}{dt}<\eta^2>.
\end{equation}
Eqs. (\ref{norm}) and (\ref{energy-dot-here}) are also useful to test
the stability of the numerical scheme we will use to simulate
Eq. (\ref{pura}).

\subsection{Newton's equations for the center of mass in a general GPE}

Let us consider the following function
\begin{equation}
  \psi({\vec r}) = \phi({\vec r}-{\vec r}_0),
\end{equation}
where $\phi$ is a solution of Eq. \ref{pura}. Substituting it in
Eq. (\ref{density}) and calculating the averaged Lagrangian, we obtain
\begin{equation}
  L = \int {\cal L} d^3r = L_{free}[\phi] + L_{cm}[\phi,{\vec r}_0].
\end{equation}

The Lagrangian has been splitted in two parts: one, the ``free''
contribution $L_{free}$, which depends only on $\phi$
\begin{eqnarray}
  L_{free}[\phi] =
  \int \frac{i\hbar}{2}\left(\phi \partial_t \phi^\ast - \phi^\ast
    \partial_t \phi\right) d^3r \nonumber \\
  + \int \frac{\hbar^2}{2m}|\nabla \phi|^2 d^3r
  + \int \frac{U_0}{2} |\phi|^4 d^3r,
\end{eqnarray}
and another one, $L_{cm}$, that includes both the potential and the
displacement ${\vec r}_0$
\begin{equation}
  L_{cm}[\phi,{\vec r}_0] =
  \int \left\{
      V({\vec r} + {\vec r}_0) |\phi({\vec r})|^2
      - i\hbar \phi^\ast \nabla \phi \dot{\vec r}_0
      \right\} d^3r.
\end{equation}

If we impose that the action be stationary for some ${\vec r}_0(t)$,
i.e. if we use Lagrange's equations (\ref{lagrange-eq}), we get
\begin{equation}
  \label{CM}
  \frac{d}{dt}{<-i \hbar \nabla>} =
  - <\frac{\partial}{\partial r_i} V({\vec r} + {\vec r}_0)>.
\end{equation}
Here the brackets denote, as usual, the mean value of an operator over the
unperturbed wave function, $\phi$, $ <A> = \int \phi^{\ast}({\vec r}) A \phi({\vec r}) d^3r.$

As a final step let us use that $\phi$ is a solution of the GPE. Then
Eq. (\ref{CM}) must be satisfied at least for ${\vec r}_0 = 0$
\begin{equation}
  \label{law2}
  \frac{d}{dt}{<-i \hbar \nabla>} =
  - <\frac{\partial}{\partial r_i} V({\vec r})>.
\end{equation}

It is now easy to prove the relation between the mean value of the
moment operator, $i \hbar \nabla$, and the speed of the center of
mass. We start from
\begin{equation}
  \frac{d}{dt}< \vec{r} > =
  \int {\vec r}(\phi^\ast \frac{\partial}{\partial t} \phi +
  \phi \frac{\partial}{\partial t} \phi^\ast) d^3r.
\end{equation}
Next we replace the time derivatives with spatial ones using Eq.
(\ref{pura}) and its complex conjugate
\begin{equation}
  \frac{d}{dt}<{\vec r}> =
  \frac{1}{i\hbar}
  \int {\vec r}(\phi^\ast \frac{-\hbar^2}{2m}\nabla^2 \phi -
  \phi \frac{-\hbar^2}{2m}\nabla^2 \phi^\ast) d^3r,
\end{equation}
and finally integrate this expression to obtain
\begin{equation}
  \label{law1}
  \frac{d}{dt}<{\vec r}> = <-i \hbar \nabla>.
\end{equation}

Eqs. (\ref{law2}) and (\ref{law1}) are the quantum equivalent of
Newton's Second Law and are exact for functions $\phi$ that satisfy the
GPE, in fact these equations coincide with the ones appearing in
the linear Schr\"odinger equation.

\subsection{The condensate in a harmonic trap. Mathieu equations for the
center of mass}

In our setup $V(\vec{r})$ is a harmonic potential with trap strengths of
the form of Eq. (\ref{lambdas}). Thus, Eqs. (\ref{law2}) become a set of
three decoupled ODE
\begin{mathletters}
\begin{eqnarray}
  \frac{d^2}{dt^2}<x> = -\frac{1}{2}m \omega^2 \lambda_x^2(t) <x>,
  \label{edo-cm-x} \\
  \frac{d^2}{dt^2}<y> = -\frac{1}{2}m \omega^2 \lambda_x^2(t) <y>,
  \label{edo-cm-y} \\
  \frac{d^2}{dt^2}<z> = -\frac{1}{2}m \omega^2 \lambda_x^2(t) <z>.
  \label{edo-cm-z}
\end{eqnarray}
\end{mathletters}
After a change of scale, all of the preceding equations are equivalent
to a model one that we will write as
\begin{equation}
  \label{mathieu-eq}
  \ddot{x} + (1 + \epsilon \cos (\omega t)) x = 0.
\end{equation}
This equation is known as Mathieu's equation. It is a well know problem
which appears frequently in the study of parametrically forced
oscillators and where one can obtain a lot of information by analytical
means \cite{Bogoliuvov,Jordan,Foale2,Thompson}.

First, Floquet's theory for linear ODE with periodic time dependent
coefficients \cite{Jordan} shows that Eq. (\ref{mathieu-eq}) has an
infinite set of instability regions in the parameter space. The limits
of these zones can be found and have the shape of wedges that start on
the points $(\omega_{min},\epsilon_{min}) = (2,0), (1,0), (2/3,0),...,$
and widen as $\epsilon$ is increased up from zero. Inside this regions
at least one branch oscillates with an exponentially increasing
amplitude.

Either with an asymptotic method, or by making use of the singular
perturbation theory, we can also locate those resonances and study the
evolution of the system around them. For a perturbation frequency close
enough to the first resonance, that is for $|\omega-2|=o(1)$, an
asymptotic method \cite{Bogoliuvov} yields up to first order
\begin{mathletters}
\begin{eqnarray}
  \sigma = \pm \sqrt{\frac{\epsilon^2}{4\omega^2} - \delta^2},
  \label{exp-fac} \\
  r \simeq c e^{\sigma t} \cos(\omega t / 2 + \theta_0).
\end{eqnarray}
\end{mathletters}
Here we see that for some values of $\delta$ and $\epsilon$ the exponent
$\sigma$ is a positive real number and the amplitude of the oscillations
grows unlimitedly. Also, the strength of the resonance is maximum for a
value of
\begin{equation}
  \label{max-eff}
  \delta_{max} = -1 + \sqrt{1 - \frac{\epsilon^2}{4}}
  \simeq - \epsilon^2 + {\cal O}(\epsilon^4).
\end{equation}
A second order Taylor expansion in Eq. (\ref{exp-fac}) lays the
following limits
\begin{equation}
  |\omega-2| \leq \frac{\epsilon}{2} + \frac{\epsilon^2}{32}.
\end{equation}

The treatment of other resonances is more difficult as they are caused
by higher order terms --at least of second order in the $\omega=1$
case--. In practice this means that they have a smaller region of
influence and that they are not so strong. One has to choose large
values of $\epsilon$, initial conditions $(x,\dot{x})$ not too close to
the equilibrium point and an excellent numerical integration method, in
order to find real instabilities. However, if one concentrates not just
in looking for exponential divergences, but on efficient pumping, it
will be easily observed that on top of these subharmonics there are
peaks of the energy gain speed (See Fig \ref{fig-spectrum-ode}b).

Finally, we wish to point out that these resonances are of a peculiarly
persistent nature: as we will precise in section \ref{losses}, they do
resist even the presence of dissipation. This has a serious and
inmediate consequence which is that feeding the condensate in a resonant
regime can result in a large amplitude oscillation of the center of
mass. On the other hand, as we will outline later in Sec. \ref{losses},
a measure of this effect can give us some insight in the losses effects,
as well as of any additional terms that could be added to Eq.
(\ref{pura}) so as to better modeling the condensate.

\section{Variational equations for the time--dependent GP equation with
  a periodic parametric perturbation}
\label{Variational}

Although the center of mass of the wave packet satisfies very simple
equations it does not happen the same to other parameters such as the
width. Only in the two dimensional case it is possible to apply moment
techniques and find analytically its time evolution
\cite{Porras,Perez95} but in the fully three dimensional problem no
exact results have been derived yet using the moment technique.

To simplify the problem, we restrict the shape of the function $\psi$ to
a convenient family of trial functions and study the time evolution of
the parameters that define that family. A natural choice, which
corresponds to the exact solution in the linear limit ($U_0 = 0$) and
provided quite good results in our previous works
\cite{theorfre,Victor2} is a three dimensional Gaussian-like function
with sixteen free parameters
\begin{equation}
  \label{ansatz}
  \psi(x,y,z,t)  =  A \prod_{\eta=x,y,z}
  \exp \left\{ \frac{-[\eta-\eta_0]^2}{2w_\eta^2}
    + i \eta \alpha_\eta+ i \eta^2\beta_\eta \right\}.
\end{equation}

For a matter of convenience and ease of interpretation we will make here
a change of parameters, from $A$ and $A^\ast$ to $N$ (the norm of the
wave function) and $\phi$ (its global phase):
\begin{equation}
  A = \frac{N}{\pi^{3/2}w_xw_yw_z} e^{i\phi}.
\end{equation}
The rest of the parameters are $w_\eta$ (width), $\alpha_\eta$
(slope), $\beta_{\eta}$ (square root of the curvature radius) and
$\eta_0$ (center of the cloud).

This trial function must now be placed in Eq. (\ref{action}) to obtain
an averaged Lagrangian per particle
\begin{eqnarray}
  \label{lagrange}
  \frac{L}{N} =
  \frac{1}{N} \int_{-\infty}^{+\infty}{\cal L}d^3r = \nonumber \\
  \hbar \dot{\phi}
  + \sum_\eta \left\{ \frac{w_\eta^2}{2} + \eta_0^2 \right\}
  \left\{ \hbar\dot{\beta}_\eta + \frac{2 \hbar^2}{m}\beta_\eta^2 +
    \frac{1}{2}m\nu^2\lambda_\eta^2(t) \right\} \nonumber \\
  + \sum_\eta \left\{ \hbar\dot{\alpha}_\eta +
    \frac{\hbar^2}{m}2\alpha_\eta\beta_\eta \right\} \nonumber \\
  + \frac{\hbar^2}{m} \sum_\eta \left\{ \frac{1}{2w_\eta^2} +
    \alpha_\eta^2 \right\}
  + \frac{U_0}{4\sqrt{2}}\frac{N}{\pi^{3/2}w_x w_y w_z}.
\end{eqnarray}

The evolution of the parameters is ruled by the corresponding set
of Lagrange equations
\begin{equation}
  \label{lagrange-eq}
  \frac{d}{dt}\left(\frac{\partial L}{\partial \dot{q_j}}\right) =
    \frac{\partial L}{\partial q_j},
\end{equation}
which give us equations for the conservation of the norm,
\begin{equation}
  \label{cons-norm}
  \frac{dN}{dt} = 0;
\end{equation}
the movement of the center of mass,
\begin{equation}
  \label{mov-center}
  \ddot{\eta}_0 + m\nu^2\lambda_\eta(t) \eta_0 = 0;
\end{equation}
the evolution of slope and curvature,
\begin{mathletters}
\begin{eqnarray}
  \beta_\eta & = & \frac{m \dot{w}_\eta}{2 \hbar w_\eta}, \\
  \alpha_\eta & = & \frac{m}{\hbar} \dot{\eta}_0 - 2\beta_\eta \eta_0;
\end{eqnarray}
\end{mathletters}
and finally the evolution of the widths,
\begin{mathletters}
  \begin{eqnarray}
    \label{widths}
    \ddot{w_x} + \nu^2\lambda_x^2(t)w_x & = &
    \frac{\hbar^2}{m^2}\frac{1}{w_x^3} +
    \frac{U_0}{2\sqrt{2}m}\frac{N}{\pi^{3/2}w_x^2w_yw_z}, \\
    \ddot{w_y} + \nu^2\lambda_y^2(t)w_y & = &
    \frac{\hbar^2}{m^2}\frac{1}{w_y^3} +
    \frac{U_0}{2\sqrt{2}m}\frac{N}{\pi^{3/2}w_xw_y^2w_z}, \\
    \ddot{w_z} + \nu^2\lambda_z^2(t)w_z & = &
    \frac{\hbar^2}{m^2}\frac{1}{w_z^3} +
    \frac{U_0}{2\sqrt{2}m}\frac{N}{\pi^{3/2}w_xw_yw_z^2}.
  \end{eqnarray}
\end{mathletters}
As one can see, the introduction of a time dependent potential
does not affect the form of the equations, which remain the same as
those of \cite{Victor2}.

Let us introduce the constants $P = \sqrt{2/\pi}Na/a_0$ (strength of the
atom-atom interaction) and $a_0 = \sqrt{\hbar/(m\nu)}$ (harmonic
potential length scale), as well as a set of rescaled variables for
time, $\tau = \nu t$, and the widths, $w_\eta = a_0 v_\eta,
(\eta=x,y,z)$. This leads us to
\begin{mathletters}
\begin{eqnarray}
\label{widths2}
  \ddot{v}_x + \lambda_x^2(t)v_x & = &
    \frac{1}{v_x^3} + \frac{P}{v_x^2v_yv_z}, \label{vx} \\
    \ddot{v}_y + \lambda_y^2(t)v_y & = &
    \frac{1}{v_y^3} + \frac{P}{v_xv_y^2v_z}, \label{vy} \\
    \ddot{v}_z + \lambda_z^2(t)v_z & = &
    \frac{1}{v_z^3} + \frac{P}{v_xv_yv_z^2}. \label{vz}
  \end{eqnarray}
\end{mathletters}
This set of rescaled variables and units is the one that we will use
throughout the paper, unless otherwise stated.

\section{Analysis of the radially symmetric case}
\label{Radial}

\subsection{The equations}

In this section we will analyse the case where the trap has the same
strength on all directions, that is
\begin{equation}
  V(x,y,z) = \frac{1}{2} m \nu^2 \lambda^2(t) (x^2 + y^2 + z^2).
\end{equation}
and the solutions are supposed to be radially symmetric.  This high
degree of symmetry simplifies the equations considerably. First, the
variational model for the widths of the condensate (\ref{widths})
reduces to a single ODE for the radial width $v(t)$
\begin{equation}
  \label{radial-ode}
  \ddot{v} = - \lambda^2(t) v + \frac{1}{v^3} + \frac{P}{v^4}.
\end{equation}
And secondly the following change of function
\begin{equation}
  \label{radial-wave}
  \psi(r,t) = A \frac{u(r)}{r},
\end{equation}
with the constrains
\begin{mathletters}
  \begin{eqnarray}
    u(r) \rightarrow 0, r \rightarrow 0 \\
    \int_0^\infty u(r) dr = 1 \\
    |A|^2 = 4 \pi N,
  \end{eqnarray}
\end{mathletters}
transforms Eq. (\ref{pura}) into the one-dimensional PDE
\begin{equation}
 \label{radial-pde}
 i \hbar \partial_t u =
 - \frac{\hbar^2}{2m} \partial_r^2 u
 + \left\{ \frac{1}{2}m \nu^2 \lambda^2(t) r^2 +
 4 \pi \frac{U_0 N}{4 \pi} \frac{|u|^2}{r^2} \right\} u.
\end{equation}
This is the equation that we have actually solved numerically.

\subsection{Numerical study of the equations}

The numerical solution of Eqs. (\ref{radial-ode}) and (\ref{radial-pde})
is not a trivial job. We can see without much effort that both equations
are stiff \cite{Hairer} when the width of the cloud becomes very small,
due to the presence of strong singular potentials. Thus, we need
numerical methods that account for the nonlinearities and are stable
enough to be trusted when close to a resonance.

To solve Eq. (\ref{radial-ode}) we have used an adaptive step size
Runge--Kutta--Fehlberg method, a Dormand--Prince pair \cite{Hairer}, the
ODE Suite of MATLAB \cite{MATLAB}, and finally Vazquez's conservative
scheme \cite{Vazquez} --a finite differences scheme that conserves a
discretized version of the energy and is unconditionally stable--. All
of them gave the same accurate results about the frequencies and
amplitude of the oscillations, the regions of divergence, etc.

To solve Eq. (\ref{radial-pde}) we have utilized a modification of a
second order accurate finite difference scheme developed in
\cite{esquema}. This new scheme is time reversible, conserves the norm
and has a discrete analogue for Eq.  (\ref{energy-dot}) which provides
enhanced stability \cite{Sanz-Serna}. On the other hand, even using the
best methods, we face another important difficulty which is the finite
size of either the spatial grid --in finite differences schemes-- or the
momentum space --in spectral or pseudospectral methods--. This size
effect becomes particularly important in the case of parametrical
perturbations and imposes a severe limit on the time for which
simulations may be trusted.

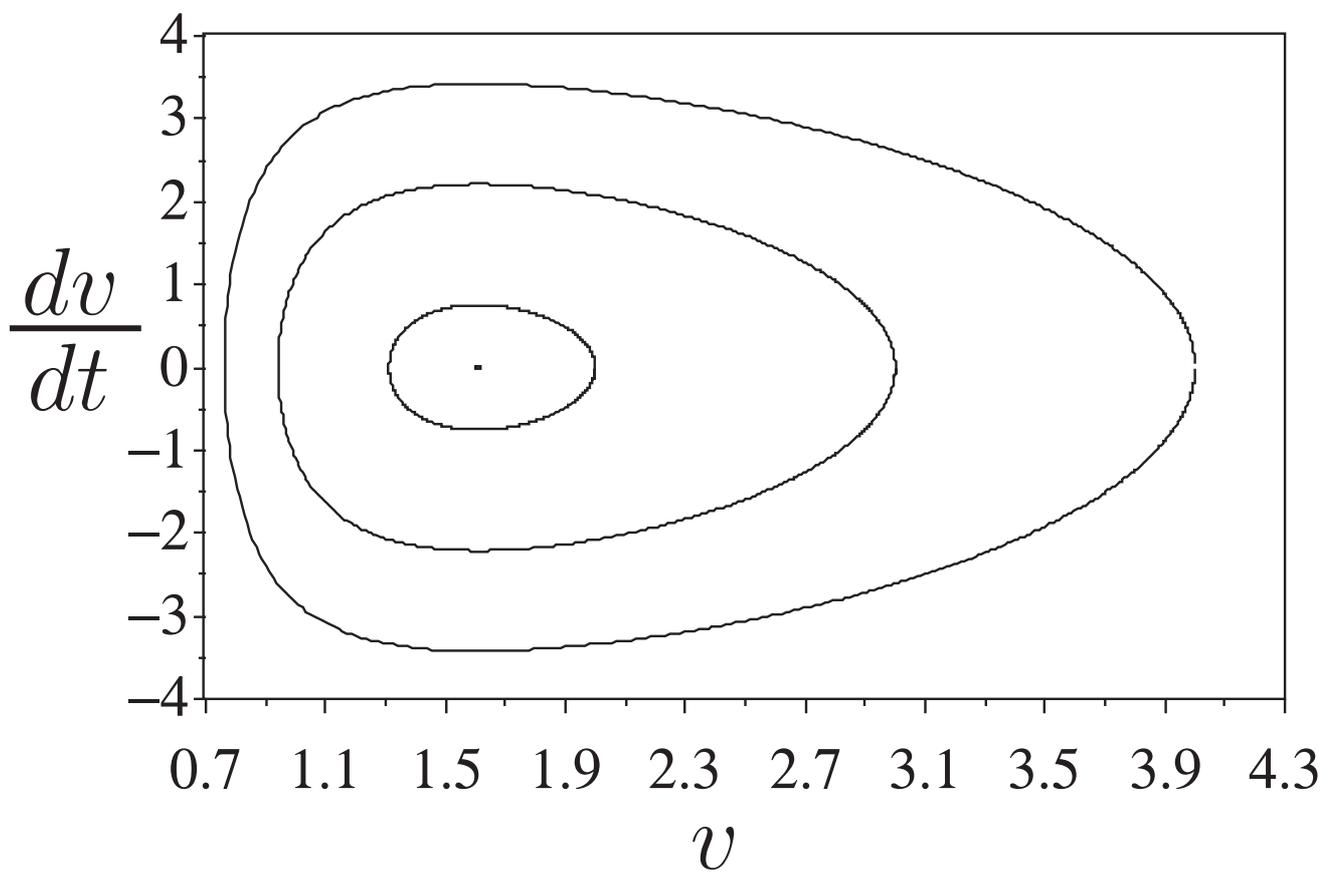
\begin{figure}
% Fig-1.ps
  \caption{ Radially symmetric condensate with $P=9.2$ as described by
    the variational equations. Phase space picture for different small
    to large amplitude oscillations.  } \label{fig-large}
\end{figure}

Using all this computational machinery we have achieved several
important results. First we have checked our programs with low
amplitude oscillations. In the PDE we imposed a gaussian initial
condition of width $v0$ and used this same value as initial condition
for Eq. (\ref{radial-ode}). By this procedure we obtained the
linearized excitation frequencies of the condensate, concluding that
there is a significant coincidence of both models as was shown in
\cite{Victor2}. In the large amplitude simulations we found a
surprising result which is that the variational model still follows
the evolution of the PDE with a 90\% of accuracy even in situations
where the condensate remains no longer gaussian but gains an important
contribution from the first and second modes. Another confirmed fact
is that the width of the condensate develops fast tough bounces
against the ``origin'' (Fig. \ref{fig-large}a-b), with its excitation
frequency deviating from the linearized predictions of \cite{Victor2}
and approaching those of the harmonic trap.

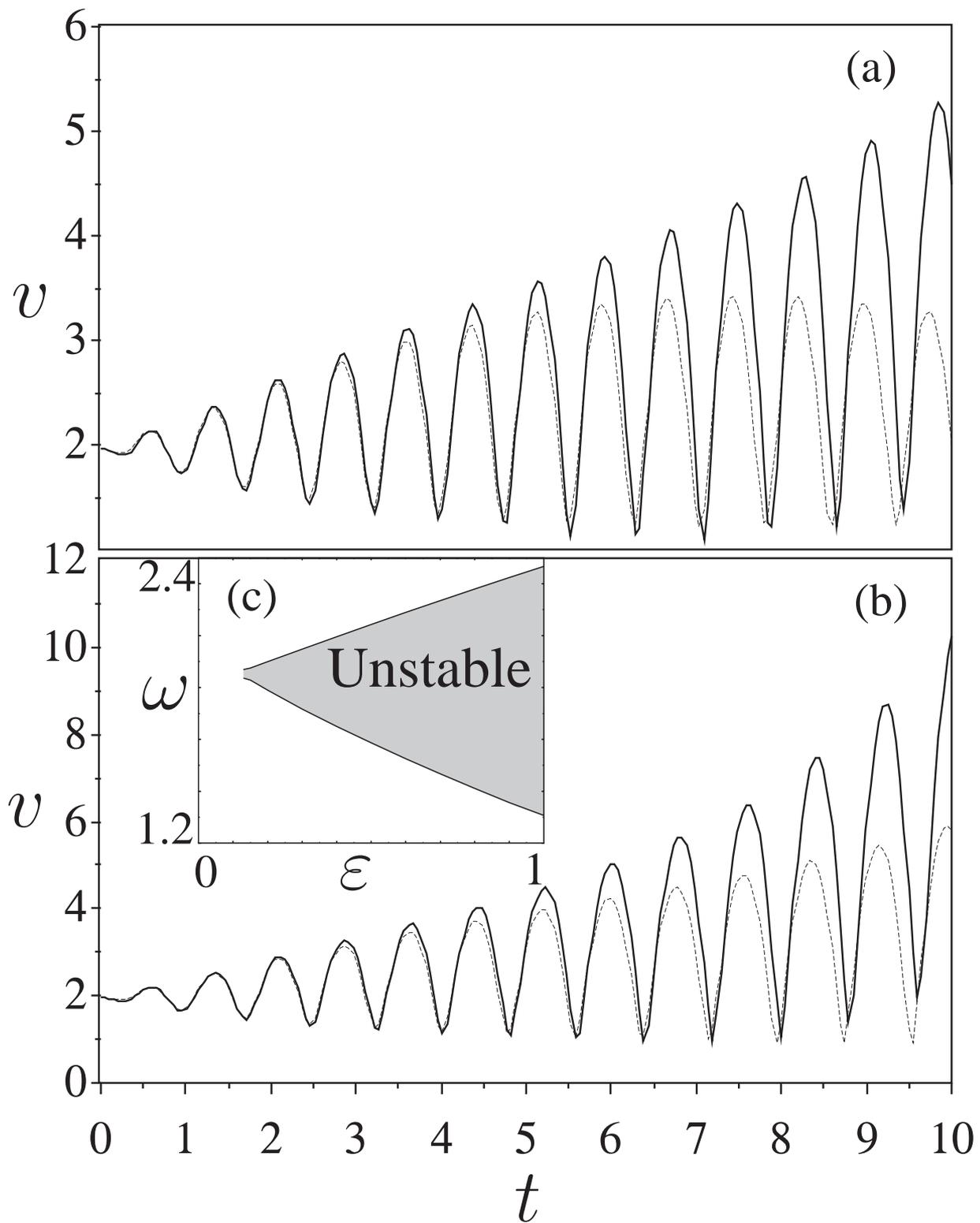
\begin{figure}
% Fig-2.ps
  \caption{ Radially symmetric condensate with $P=9.2$ and initial
    condition $v=1.6$, subject to a periodic perturbation. Evolution of
    the width in the variational model (dashed lines) and in the GPE
    (solid line) when the perturbation is (a) $(\epsilon = 0.15,
    \omega = 4.00)$ and (b) $(\epsilon = 0.2, \omega = 4.00)$. (c)
    Instability region for the variational model in the parameter
    space around the main resonance.}
\label{fig-divergence}
\end{figure}

We have also simulated the system with a periodic time dependent
perturbation of the form
\begin{equation}
  \label{radial-lambda}
  \lambda^2(t) = 1 + \epsilon \cos (\omega t).
\end{equation}
Using the variational equation Eq. \ref{radial-ode} we have scanned the
parameter space $(\omega,\epsilon)$. We have found one region where the
radial width diverges exponentially (Fig. \ref{fig-divergence}a) and
causes most numerical methods to fail after a finite time, and two more
where the width grows almost exponentially up to a point where the
simulation cannot account for this growth. All of these zones have the
form of wedges, with a peak at $(\omega_{min}, \epsilon_{min})$, and a
growing width as $\epsilon$ is increased. The most important resonance
stands on $\omega_{min} = 2.04$ (See Fig. \ref{fig-divergence}b). The
other ones are weaker and rest on $\omega_{min} = 1.02, 0.68$.  We have
studied a wide range of setups and found that these frequencies change
no more than a $0.5\%$ depending on the initial conditions and the
nonlinearity. On the other hand, the lowest perturbation amplitude for
which the resonance exists $\epsilon_{min}$ does exhibit a strong
dependence on the initial conditions. Indeed, around the weaker
resonances, instability is never reached for points close to the minimum
of the potential. However, as we already pointed in Sec. \ref{Center}
this is probably a numerical effect.

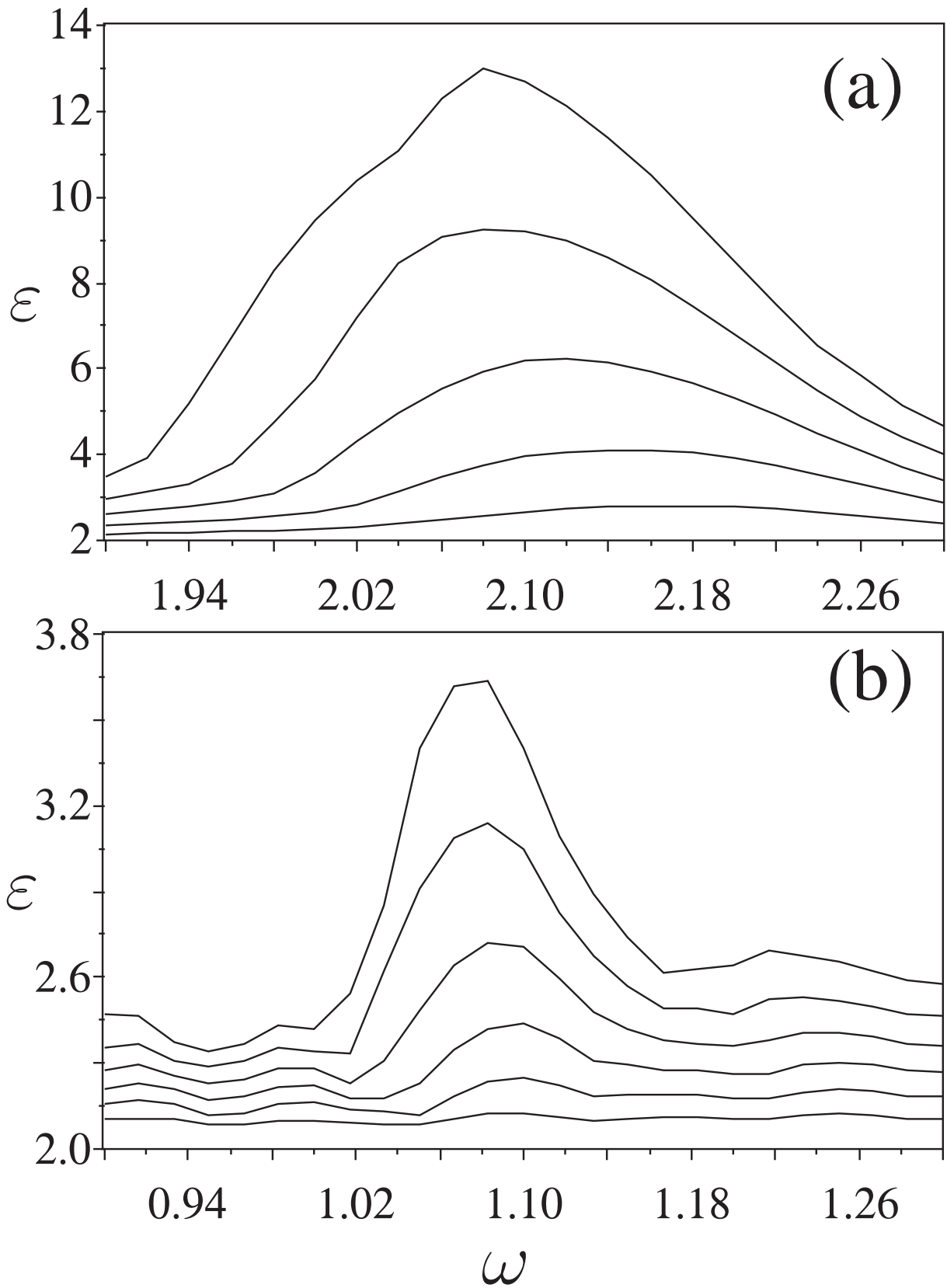
\begin{figure}
% Fig-3.ps
  \caption{ Radially symmetric condensate with $P=9.2$. Plot of the
    maximum amplitude of oscillations for the GPE after 40 time
    units. The initial condition corresponds to gaussian of width
    $v=1.6$. Each line corresponds to a different value of $\epsilon$,
    from 0.05 to 0.3 in steps of 0.05. The frequency range covers (a)
    the main resonance and (b) the second important one.}
  \label{fig-spectrum-pde}
\end{figure}

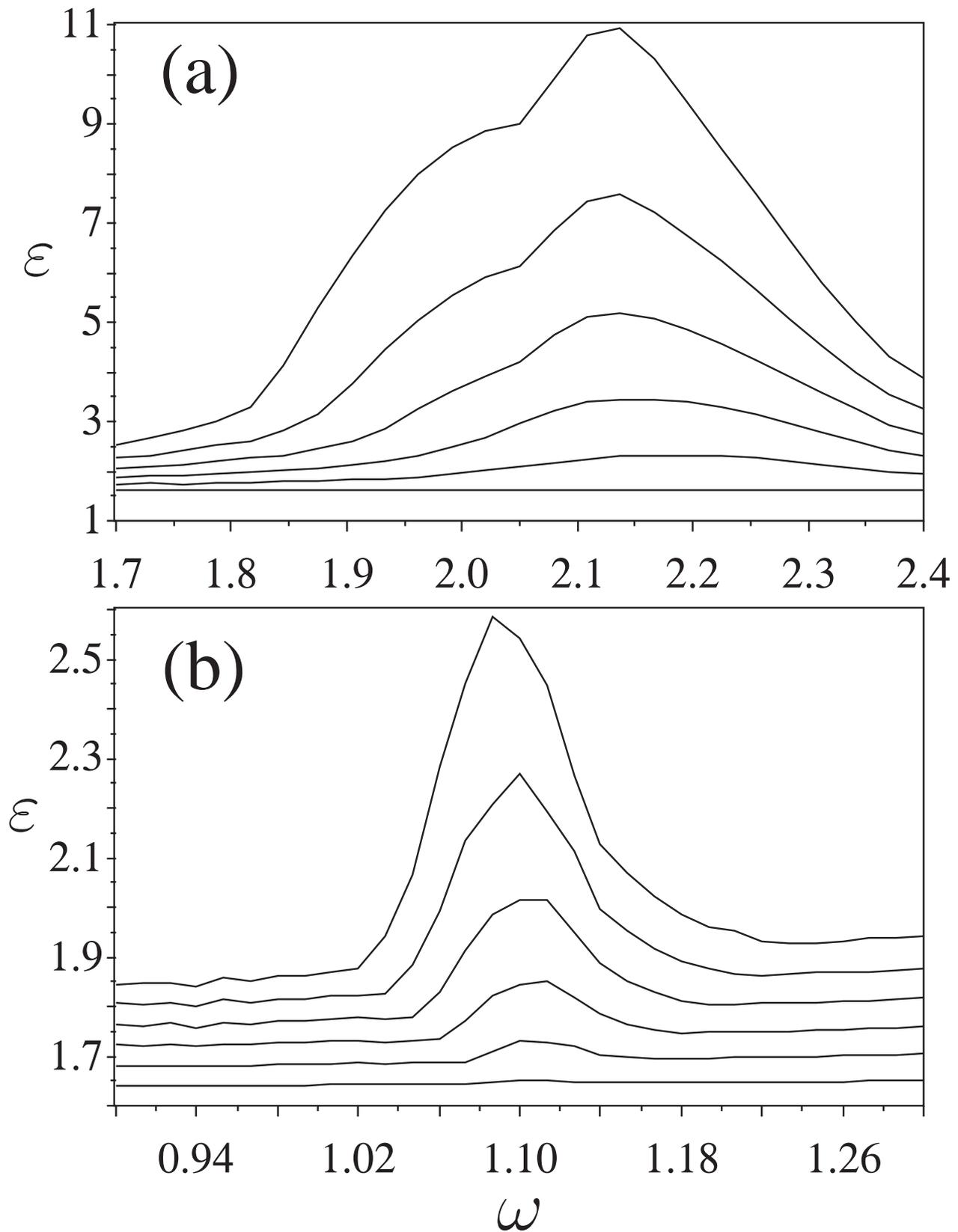
\begin{figure}
% Fig-4.ps
  \caption{ Radially symmetric condensate with $P=9.2$. Plot of the
    maximum amplitude of oscillations for the variational model after 40
    time units. Initial conditions are $v=1.6, \dot{v}=0$. Each line
    corresponds to a different value of $\epsilon$, from 0.05 to 0.3 in
    steps of 0.05. The frequency range covers (a) the main resonance and
    (b) the second important one.}
  \label{fig-spectrum-ode}
\end{figure}

In order to study the location of the resonances we have also made some
plots of the efficiency of the energy absorption process against the
perturbation frequency for different values of the perturbation
amplitude for the variational system and the partial differential
equation (Figs. \ref{fig-spectrum-ode}a-b,
\ref{fig-spectrum-pde}a-b). The way we have measured ``efficiency'' is
by letting the system evolve for a fixed time and then computing the
maximum value of the mean square radius of the cloud. For the sake of
simplicity and to approach the experimental setups, we have used an
equilibrium state of the static GPE as initial condition.

In a rather complete inspection of the parameter space using the
efficiency plots, we have found that, though the resonance regions
cannot be precisely delimited because of the dependence on the initial
data, they do exist and behave much like the variational model
predicted. As we see in Figs. \ref{fig-spectrum-ode} and
\ref{fig-spectrum-pde} there are two important features in the response
of the condensate. The main one is the width of the resonances and the
dependence of that width on the strength of the perturbation. The second
important feature is the change of the frequency for which the
perturbation is most efficient. This peak is centered on the frequency
of the linearized model only for very small perturbation amplitudes, and
switches to the trap natural frequencies very quickly as the amplitude
is made stronger -- of about $10\%$ or so--. For even stronger
amplitudes the optimal frequency decreases slowly.

Another consequence of this work is that response of the cloud is
stronger in the PDE than in the variational simplification. For
instance, it is possible to find (See Fig. \ref{fig-divergence}a)
perturbation amplitudes that do not cause an significant growth in
Eq. (\ref{radial-ode}) but make the cloud width increase quite linearly
in the PDE. We have also studied the case where an exponential growth
of the width is present in both equations --(\ref{radial-ode}) and
(\ref{radial-pde})-- and we have seen that the growth is qualitatively
similar though there is a tendency in the exact model to exhibit
slightly larger amplitude motion. These discrepancies are originated by
high modes which are not present in the variational treatment. In fact,
even though an important part of the cloud remains close to the origin,
it is observed that for long times a long tail appears which is hard to
appreciate but is responsible for the growth of the mean square width
(and the presence of higher order modes).

\subsection{Analysis of the Mathieu equation with a singular potential}

In this subsection we are to develop a simplified model that explains
why do resonances appear in the perturbed GPE equation. This model makes
use of the variational equations for the cloud width but {\em does not
care for the actual shape of the variational ansatz}. The reason for it
is that both the approximate model and the exact one agree for short
times, the response of the latter being always stronger.

Let us limit ourselves to Eq. (\ref{radial-ode}). In the previous
subsection we said that this system is very stiff and that the origin
acts as an elastic wall. In view of this, it is intuitively appealing
to replace the singular but differentiable potential $1/v^3 + P/v^4$
with a discontinuous bounce condition on the origin, i.e. an impact
oscillator. Numerical simulations confirm that this approximation is
good for large amplitude oscillations. Also, the connection between
soft singular potentials and impact oscillators has been widely
studied and these kind of models have found ample application in
real--life situations in the fields of Physics and Engineering (See
\cite{Foale2,Thompson} for many references).

Since the variational model is a lossless one, our boundary condition
must be elastic. We replace Eq. (\ref{radial-ode}) with the following
one
\begin{eqnarray}
  \label{bounce}
  \ddot{v} + \lambda^2(t) v & = & 0, \\
  \lim_{t \rightarrow t_c^-}(v,\dot{v}) = (0^+, V_c)
  & \iff & \lim_{t \rightarrow t_c^+} (v,\dot{v}) = (0^+, -V_c), \nonumber
\end{eqnarray}
where $t_c$ denotes any isolated instance when the system bounces
against the $v=0$ singularity.

Let us show that this equation is in turn equivalent to an elastic
oscillator {\em without} barrier conditions. We introduce the change of
variables
\begin{equation}
\label{change-barrier}
 v = |u|,
\end{equation}
where $u$ is an unrestricted real number and satisfies the following
one--dimensional harmonic oscillator equation
\begin{equation}
\label{harmonic-osc}
\ddot{u} + \lambda^2(t) u = 0,
\end{equation}
It is now easy to prove that every solution of Eq. (\ref{harmonic-osc})
provides a solution of Eq. (\ref{bounce}). And vice versa, from every
solution of Eq. (\ref{bounce}) it is possible to construct a solution of
Eq. (\ref{harmonic-osc}), unique up to a sign.

So, what do we have now? We have proved that for large amplitude motion
Eq. (\ref{radial-ode}) behaves as Mathieu's equation. This implies that
in the regime of medium to large amplitude oscillations, or in a
situation of large amplitude perturbations, Eq. (\ref{radial-ode}) will
have instability regions that are more or less centered on Mathieu's
frequencies $\omega = 2, 1, 1/3,...$. This prediction is indeed
confirmed by the numerical simulations: the main resonance is capable of
causing an exponential divergence, while the other ones are harder to
track down but do appear in plots of pumping ``efficiency''.

Another, important but minor result of this equivalence is that these
resonance regions must become wider and move on to smaller frequencies
as the perturbation amplitude is increased. This result is also obtained
in the numerical simulations. Indeed, from the graphs we can estimate
how fast the optimal frequency decreases as $\epsilon$ grows, and we
will see that the order of magnitude corresponds to that of
Eq. (\ref{max-eff}).

Summing up, what we have seen here is that for low amplitude
oscillations the condensate moves in an effective potential that is
parabolic. Thus, the frequency that excites the condensate most
efficiently in a parametric way is the one that results from the
linearization of Eq. (\ref{radial-ode}). On the other hand, when we
start to consider large amplitude oscillations, we find that the
harmonic trap gains importance over the details of the well. It is in
this situation that the instability arises, and it happens for
perturbations that oscillate according to multiples of the frequency
of the trap.

Even more, higher modes are little influenced by the nonlinearity, just
because they are more spread and the value of $|\psi(r,t)|^2$ is
smaller. As a consequence, the energies of these modes come even closer
to those of the linear harmonic oscillator, resulting in the fact that
the response of the GPE is stronger than what the variational model,
limited to a gaussian shape, predicts.

\section{Analysis of the nonsymmetric case}
\label{Nonsym}

After finishing the study of the radially symmetric problem, it seems a
natural step to proceed with the nonsymmetric one. However this step is
not simple for several reasons, the main one being that the simulation
of full three dimensional GPE is computationally very expensive work.
Thus, it would be unwise to directly attack the full problem without
gaining some insight on what is to be expected from the simulations by
cheaper means.

In our case the cheapest tool is the variational model. We have seen
that it describes rather well the behaviour of a radially symmetric
condensate. And, as we pointed out before, there are exact analytical
studies \cite{Porras,Perez95} where the moment method reduce exactly
the {\it two dimensional} GPE to a set of ODE which are similar to the
ones we have.

\subsection{Predictions of the variational model}

When we remove the radial symmetry in the variational equations, we are
left with two to three coordinates, and the perturbation can bear many
different forms. However, trying to follow the experimental setups
\cite{JILA}, we should once more take a sinusoidal time dependence for
every $\lambda_{\eta}(t)$ coefficient, as in Eq. (\ref{lambdas}). This
choice accounts both for the $m = 0$ mode $(\epsilon_x = \epsilon_y,
\epsilon_z = 0)$ and the $m = 2$ perturbations $(\epsilon_x =
-\epsilon_y, \epsilon_z = 0)$ from the JILA experiment \cite{JILA}. In
the latter case the potential is a parabolic one, with fixed frequencies
on a rotating frame. However, for our trial function (Eq. (\ref{pura})
it behaves just as a potential of the form of Eq. (\ref{parabolic})).

Substituting our effective perturbation frequencies into Eq.
(\ref{widths2}) we get a set of three coupled Mathieu equations with a
potential that is singular on the $v_x = 0$, $v_y = 0$ and $v_z = 0$
planes. The singularities are at least as strong as $1/v^3$, and the
numerical simulations again confirm that they act as elastic walls, so
we now proceed with a change of variables formally equivalent to that of
Eq. (\ref{change-barrier}):
\begin{mathletters}
  \label{decoupled-3d}
  \begin{eqnarray}
    w_\eta & = & |u_\eta| \\
    \ddot{u}_\eta & = & -\lambda_\eta^2(t) u_\eta
  \end{eqnarray}
\end{mathletters}
for $ \eta=x,y,z$.

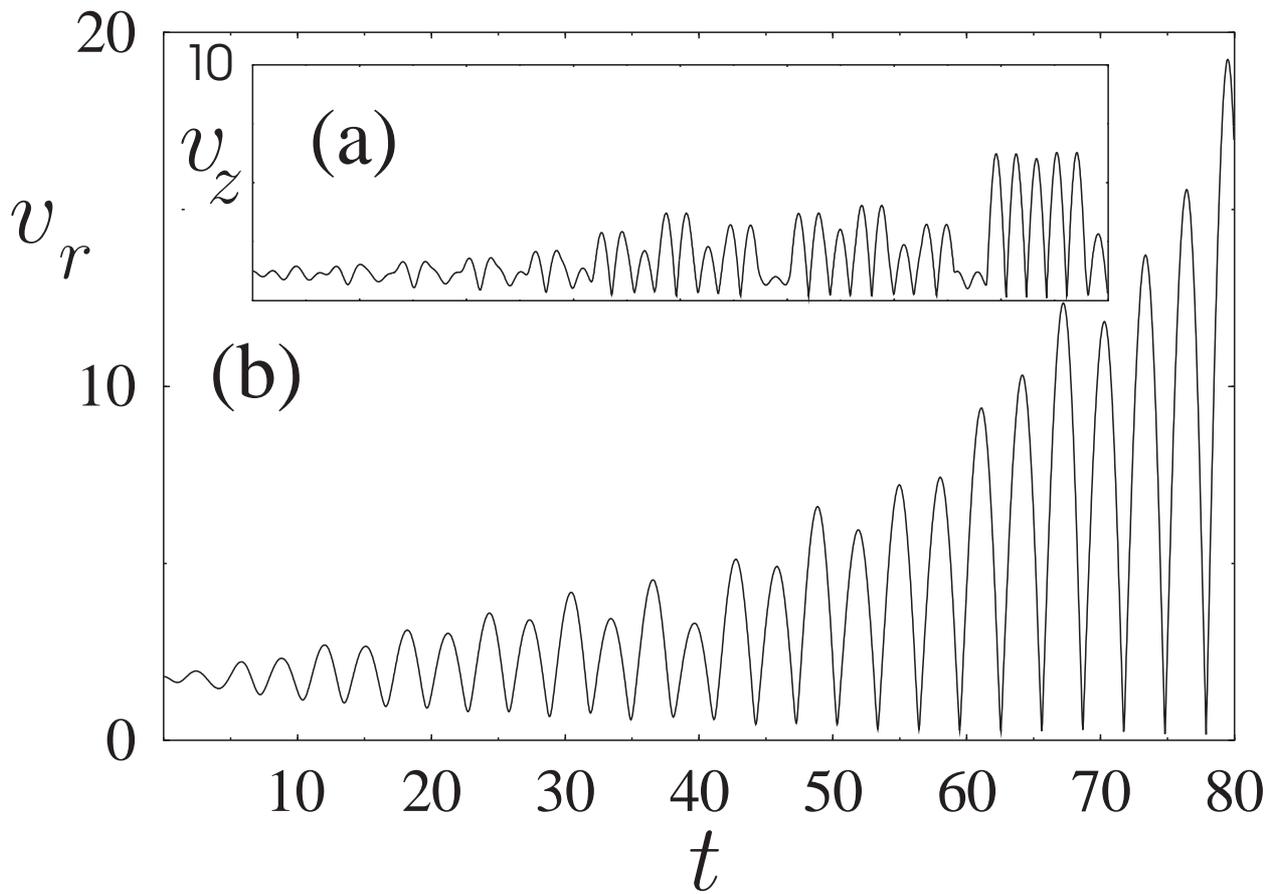
\begin{figure}
% Fig-5.ps
  \caption{ Evolution of a cylindrically symmetric condensate
    (variational model, $P=9.2$) under a sinusoidal perturbation
    $(\omega,\epsilon)=(2.04,0.1)$ of only the radial strength of the
    trap.  Both (a) the radial $v_r$ and (b) the axial $v_z$ widths are
    plotted.}
  \label{fig-cylindric}
\end{figure}

Now the situation is a bit more complex. The first new feature is the
existence of several sets of instability regions. Due to having three
{\em a priori} different constants $\lambda_{0\eta}$, the three
oscillators in Eq. (\ref{decoupled-3d}) are not equivalent and we may
get three sets of resonances in the $(\epsilon_\eta, \omega)$ space,
each one containing the instability regions that start on the
frequencies
\begin{equation}
  \frac{\omega_{\eta,min}}{\lambda_{0\eta}} = 2, 1, 2/3, ...
\end{equation}
Numerical simulations of the variational equations for the $m=0$ type
excitation confirm this prediction with a relative accuracy around
$0.5\%$ in the frequencies. The results show again that, opposite to
the pure Mathieu equation, these ``wedges'' rest on a nonzero value of
the perturbation amplitude $\epsilon_{min,\eta}$. In
Fig. \ref{fig-cylindric} we see the evolution of the condensate width
with parameters close to the main resonance region.

Another new feature is the possibility of coupling between the widths of
the condensate. This coupling is seen both in the $m=0$ and $m=2$
excitations setups. In the first one, the perturbed width feeds the
unperturbed one. Figure \ref{fig-spectrum-pde} demonstrates that
efficiency is not very high. In the $m=2$ case two widths are associated
to the same trap frequencies and the perturbations are of equal
magnitude and opposite sense. The explanation is that both widths block
each other, eliminating the resonance and leaving just a bounded
``movement'' of small amplitude.

As a side note, we must say that the variational model itself predicts
the lack of resonances in the $m=2$ setup. We already pointed out that
the JILA $m=2$ perturbation corresponds to a situation where the
potential of the trap is rotated while maintaining its shape. It can
be easily proved that if we choose {\em any} family of trial functions
with enough degrees of freedom for a general rotation, the variational
solution will always stick to the potential and rotate at the same
speed. This also implies that the trial function (\ref{ansatz}) is not
suitable for describing the condensate when this kind of perturbation
is applied.

\subsection{Numerical study of the three dimensional GPE}

To perform the simulation of the full GPE we have used a fourier
pseudospectral method, using typically a grid of $108^3$ collocation
points and integrating in time with a second order, symmetrized split
step method \cite{bpm1,bpm2}.

Our numerical study is based on a $O((\Delta t)^2)$ scheme which behaves
extremely well for long time runs. However, as in the radial PDE, no
matter how accurate the scheme is, there's a limit in the time during
which simulations can be trusted and this limit is imposed by the growth
rate of the condensate width and the size of the grid. This limit is
specially important for the $m=0$ perturbation, where the condensate
develops a large tail in the unperturbed direction, and this tail breaks
the simulation after a certain time. The more asymmetric the trap is,
the sooner this effect comes out. In our simulations the grid was a box
of $108^3$ equally spaced points and whose sides measured from 20 to 40
length units, depending on the symmetry and intensity of the trap. This
allowed us to track the condensate for about 12 periods in truly
resonant setups, and for much more in nonresonant ones.

We have applied the algorithm to many different problems. First we
checked our programs against stationary and radially symmetric problems.
Secondly we introduced time dependent traps and reproduced the
calculations of Section \ref{Radial}. In both cases we got the expected
results.

The third set of experiments consisted in a resonant time dependent
radially symmetric trap applied to several slightly asymmetric gaussian
wave packets. The initial asymmetry was slight enought to treat it as a
weak perturbation and we saw that it departured little from the
symmetric case --i.e., no modes with higher energy or angular moments
break the exponential growth.

The fourth and probably most important group of simulations consisted in
the study of the $m = 0$ perturbation from the JILA \cite{JILA}
experiment. Here we confirmed the predictions of the previous
subsection, that is we obtained at least one resonance region where the
variational model predicts.

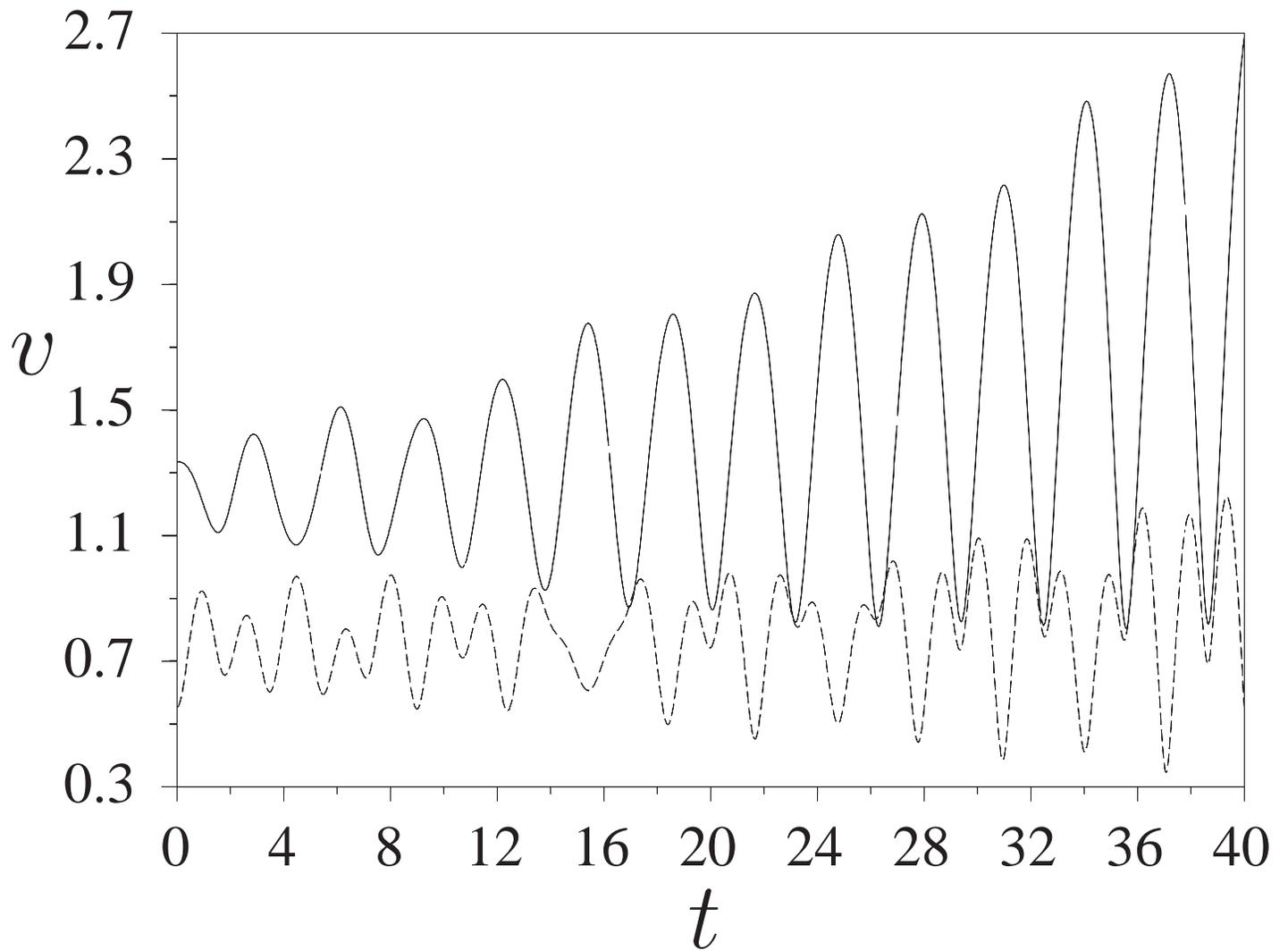
\begin{figure}
% Fig-7.ps
  \caption{ Evolution of a condensate in a cylindrically symmetric trap
    ($P=9.2, \lambda_r=1, \lambda_z=2$) subject to a $m=0$ perturbation
    $(\omega_r,\epsilon_r)=(2.00,0.15)$. Both the radial width $v_r$
    (solid line) and the axial $v_z$ width (dashed line) are plotted.}
    \label{fig-cylindric-pde2}
\end{figure}

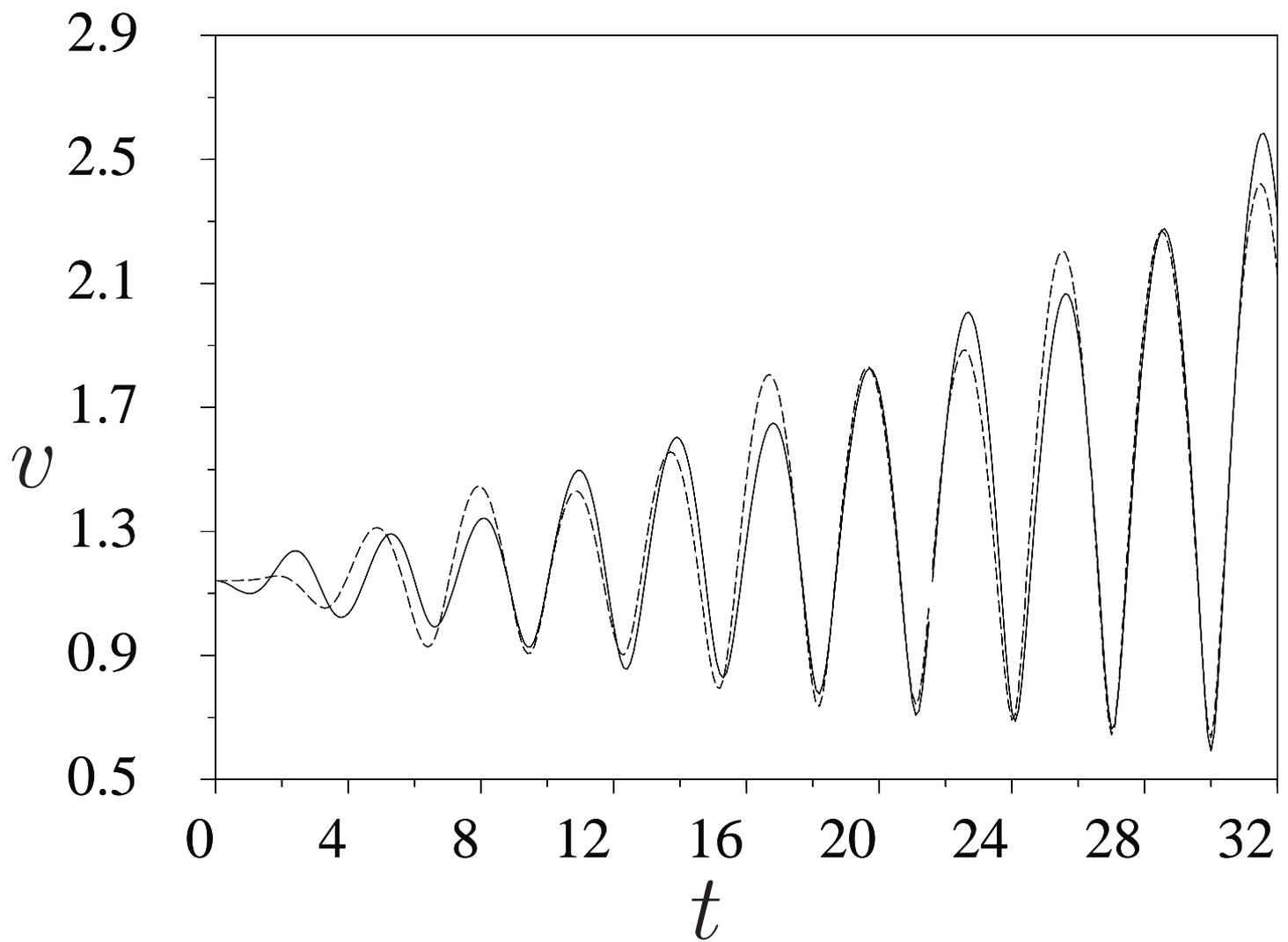
\begin{figure}
% Fig-8.ps
  \caption{ Evolution of a condensate in a spherically symmetric trap
    ($P=9.2$) subject to a $m=0$ cylindrical perturbation
    $(\omega_r,\epsilon_r)=(2.00,0.15)$.  Both the radial width $v_r$
    (solid linea) and the axial width $v_z$ (dashed line) are plotted.}
    \label{fig-cylindric-pde1}
\end{figure}

We also observed that the response of the condensate as modeled by the
GPE is stronger and exhibits a slightly more intense growth rate than
what the variational model predicts. This and other similar results from
the radial case (Sect. \ref{Radial}) favor the theory of a cooperative
staircase effect, where the higher modes contribute to the energy
absorption process without interfering in the evolution of the
width. Figs.  \ref{fig-cylindric-pde1} and \ref{fig-cylindric-pde2} show
just two examples of the kind of evolution of have seen for this
perturbation model on the resonant regions.

The reason for this cooperative effect seems to lay in the energy level
structure of the GPE equation. We have studied the evolution of the
correlation a condensate wave against its initial data. In all cases the
initial data was a displaced and deformed gaussian cloud, while the
environment corresponded to a stationary trap. In the linear case it is
easy to show that the spectrum of the correlation must reveal a subset
of the eigenvalues of the hamiltonian, and indeed that is what we got.
In a nonlinear context it is not clear what the frequencies of the
correlation mean --they may be or may be not eigenvalues of the GPE--,
but at least we know that they must rule the energy absorption process
somehow. What numerical experiments show is that for an extensive family
of initial conditions these generalized ``spectra'' can be approximated
by the formula $E_n = \omega n + E_0$, where $E_0$ depends on the
nonlinearity and the level spacing is a regular, harmonic one.

This picture of equally spaced levels is intuitively appealing to
explain the existence of such strong resonances. On one side we have the
fact that, in any other system, a continous parametric perturbation
would become bounded as the higher state become populated. This is
probably what one would first think when facing this system. On the
other side we see that due to being equally spaced these higher energy
states are themselves sensitive to the same resonant frequency and offer
no resistence to the particle promotion process. In the end, this is
more or less what happens in the variational equations, considered from
the point of view of Classical Mechanics.

However, it is not relevant for the existence of resonances whether they
cause a sustained growth or not. What is confirmed without any doubt is
that the perturbations that are most efficient in the variational model
are also the most efficient in the full GPE.

\section{Analysis of the effect of losses}
\label{losses}

We now want to show the effect of a dissipative term in the variational
equations. This term will be introduced in a phenomenological way so as
to model the damping of the oscillations of the condensate in regimes
where the number of noncondensed particles is small. We will choose a
viscous damping term that models well the behaviour of the condensate in
the experiments \cite{JILA}. This choice introduces a significant
loss of energy in the oscillations while preserving the number of condensed
particles. Using this term we will see that the resonance regions of the
pure Mathieu persist.

Let us see what Eq. (\ref{mathieu-eq}) looks like once we add damping:
\begin{equation}
 \label{mathieu-disip}
 \ddot{x} + (1 + \epsilon \cos (\omega t)) x + \gamma \dot{x} = 0.
\end{equation}
A simple change of variables $x(t) = \rho(t) e^{-\gamma t}$ makes this
new term disappear, transforming it back to a pure Mathieu equation
\begin{equation}
  \label{damped}
  \ddot{\rho} + (1 - \gamma^2 + \epsilon \cos (\omega t)) \rho = 0.
\end{equation}
With the introduction of the damping we are shifting the resonances to
lower values, given by
\begin{equation}
  \frac{\omega}{\nu(\gamma)} = 2,1,2/3...
\end{equation}
where $\nu^2(\gamma) = 1 - \gamma^2$ is the new effective frequency for
the trap. We can approximately solve Eq. (\ref{damped}) around the
first resonance, obtaining
\begin{equation}
  x(t) \simeq c e^{(\sigma - \gamma)t} \cos(\frac{\omega t}{2} +
  \theta_0),
\end{equation}
where $\lambda$ is given by Eq. (\ref{exp-fac}).

This shows that the resonance regions in the parameter space are
constrained to values of $(\omega, \epsilon)$ for which the strength of
the resonance, $\lambda$, is larger that the strength of the dissipative
term. The new regions have a larger, nonzero, value of $\epsilon_{min}$,
and are typically thinner, but do not disappear unless $\gamma$ is very
large.

This effect is reproduced in the variational model when we introduce
similar viscous damping terms. For instance, taking the data from the
JILA experiment \cite{JILA}, we can estimate a condensate lifetime of
about 110 ms and a value for $\gamma$ of $0.15$ in natural units of the
condensate. Such damping makes the $\epsilon_{min}$ value raise from
$0.09$ to $0.18$ for the $P=9.2$ case. Thus, the instability should not
be appreciated unless the perturbation amplitude exceeds the 20$\%$.

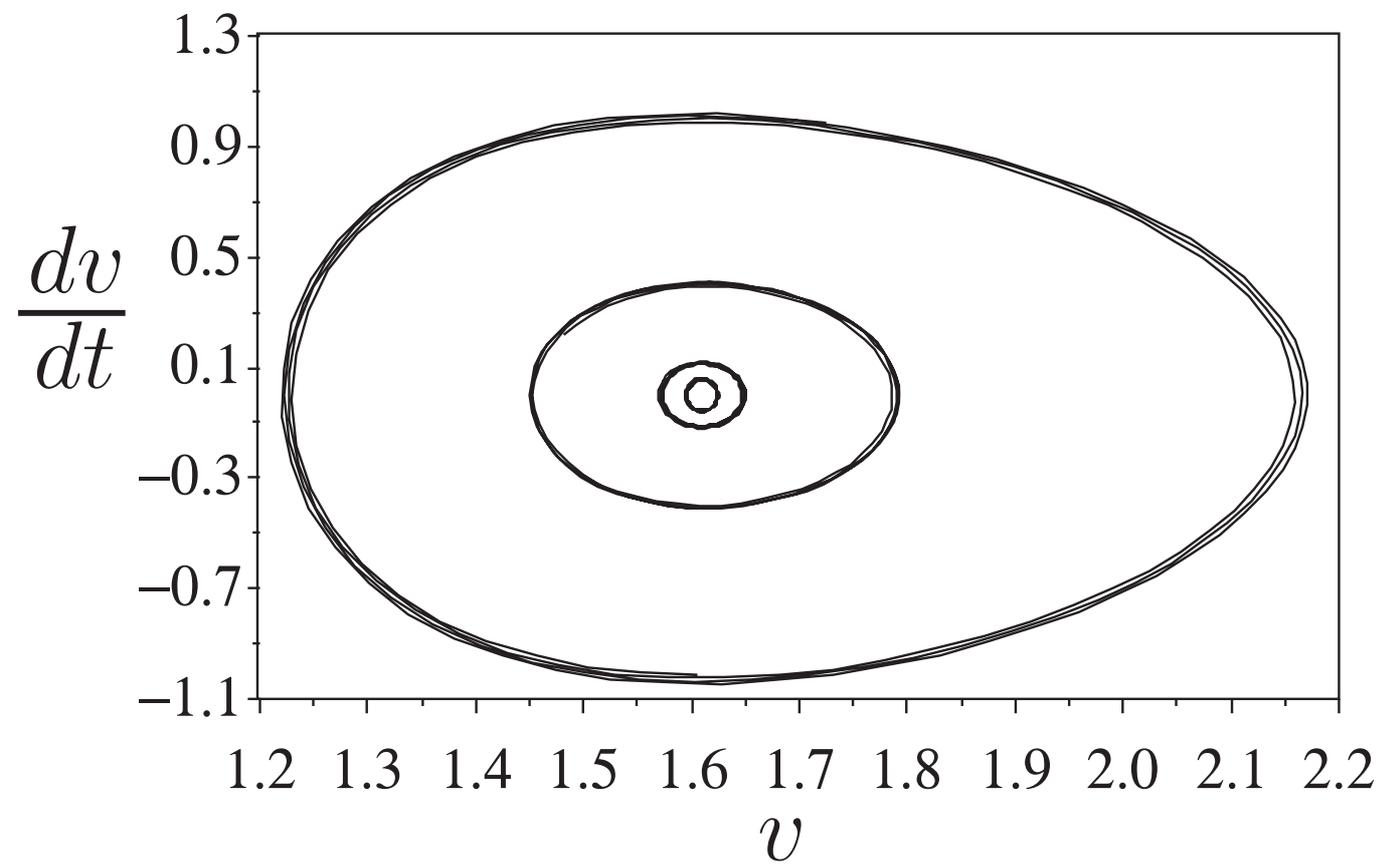
\begin{figure}
% Fig-6.ps
  \caption{ Phase space picture for the width of radially symmetric
    condensate subject to damping plus a periodic perturbation. The
    nonlinearity is $P=9.2$, the damping $\gamma=0.15$. The perturbation
    has in all cases $\epsilon=0.1$, while the frequencies are,
    from the outer cycle to the inner one, $\omega=2.15, 2.4, 3.0, 4.0$. }
  \label{fig-damped}
\end{figure}

An interesting effect of damping is that the evolution of a continuously
perturbed condensate outside the instability regions becomes more
ordered than in the undamped mode since the motion is constrained to
limit cycle {\em synchronized to the frequency of the parametric
perturbation}, and with a size that depends only on the perturbation
parameters, $(\omega,\epsilon)$. In Fig. \ref{fig-damped} we show the
different limit cycles that appear under peridic perturbations. The
largest one is always the one with its frequency on top of the peak of
the resonance as shown by Fig. \ref{fig-spectrum-pde} and the size of
the limit cycle decreases as the frequency is detuned from this value.

Finally we wish to point out that the appearance of a limit cycle opens
the door to a wide family of phenomena, from chaotic motion to
bifurcation theory \cite{Thompson2,Thompson3}. This limit cycle would
exist under a great variety of dissipative terms, and is not exclusive
of linear damping. Also, the dependence of the limit cycle on the
damping constant can be useful from the experimental viewpoint to
separate the condensed and noncondensed clouds as will be pointed out in
the last section.

\section{Conclusion and discussion.}
\label{Conclu}

In this work we have analysed the resonant dynamics of the
parametrically forced time dependent GPE using exact analytical
techniques, approximate time dependent variational techniques and
numerical simulations. All the results point to the existence of various
resonant behaviors associated to the same parameter regions concerning
the motion of the center of mass and the width oscillations of the wave
packet. The role played by the nonlinearity is to provide a strong
repulsive term at the origin which acts as a barrier, which depending on
the dimensionality and the symmetry of the external forcing could be
stronger than the repulsive term related to the linear dispersion
(kinetic energy term).

We have developed a simplified version of the variational equations for
the spherically symmetric condensate under periodic sinusoidal change of
the trapping potential. This model is based on an impact oscillator,
i.e. a harmonic oscillator with an elastic barrier condition which
allows to find explicit solutions.  When applied to the parametrically
perturbed condensate, our model predicts that medium to large
oscillations approach the harmonic trap frequencies, not the ones
resulting from the linearization of the variational equations. The model
also predicts that the response of the condensate to an external
perturbation is ruled by the harmonic trap frequencies. In the case of a
sinusoidal parametric perturbation it accurately predicts the existence
of a family of resonances on multiples of the trap natural frequencies.

In the end, what we get from this work is a precise picture of the
resonances for a wide family of equations that include the Eq.
(\ref{pura}) and the harmonic oscillator. In this description for the
radially symmetric case we have one base frequency that is essentially
the same for both equations (linear and nonlinear one) and which
corresponds to the energy separation between the ground state and the
first excited state of the harmonic oscillator up to a high precision.
The invariance of this base frequency has been checked for a
nonlinearity constant going from $P=9.2$ --the JILA \cite{JILA}
experiment-- to 20 times this value. It differs from the predictions of
\cite{Yukalov}, where formulas regarding the $P \rightarrow 0$ and $P
\rightarrow \infty$ are derived. Also there is a whole set of
subharmonics of this frequency, all of which are capable of exciting the
cloud quite efficiently. At least three of them have been found, both
with significant responses. These subharmonics are not predicted in
\cite{Yukalov}.

We have also found that for this kind of parametric drive the resonances
are wide. The width grows with the strength of the interaction and
decreases with the effect of dissipation. Both facts can be checked in
the experiments by forcing the system for a longer time than what it is
currently done.

Both predictions are exact in the linear limit, $U_0=0$, and have been
confirmed with simulations coming both from the exact variational model
and the GPE. The discretization of the GPE has also allowed us to scan
the parameters space, studying the efficiency of the perturbation
process. In this study we have only found peaks centered on Mathieu's
frequencies.

For the general case with more than one degree of freedom (axial and
nonsymmetric cases) we obtain a set of two to three decoupled pure
Mathieu equations. We have shown that, due to having more than one
frequency, the predicted Mathieu resonances do exist in a larger number.
On the other hand, we have also seen that some of these resonances may
disappear due to the locking of ``equivalent'' variables, an effect that
our decoupled equations do not account for.

This resonance scheme for the nonsymmetric case has been confirmed with
accurate simulations of the full GPE for the $m=0$ perturbation. An
analysis of the correlation of a state against its initial data shows
that both the linear and the nonlinear problems exhibit a spectral
structure which is likely to present such behavior.

Finally, damping has been shown to limit the effect of the parametric
perturbation.  Once more we have proved that only frequencies close to
the Mathieu resonance regions do excite the condensate as a whole in an
efficient way, causing the appearance of a stable limit cycle. All other
frequencies are inefficient in the sense that the system stays {\em
extremely} close to the equilibrium configuration, which acts as a
focus.

A main conclusion of this work is that for this set of resonances to
exist one only needs a singularity that prevents from collapse. The
variational method showed that the kinetic terms in the evolution
equations guarantee a $1/x^3$ singularity as far as we impose a
repulsive interaction between the atoms in the cloud. This is the reason
why we say that we have a family of systems that behave much the same.
An immediate result of this is that the response of the noncondensed
atoms under the parametrical perturbation will be qualitatively similar
to that of the condensed ones, with the only difference that the former
are subject to a more intense dissipation.  But as we already saw in
Sec. \ref{losses} this dissipation can be enough to distinguish both
kind of fluids: while the condensed part might suffer an exponential
growth, the uncondensed part might develop low amplitude bounded
oscillations.

We have also demonstrated that these resonances show up in the movement
of the center of mass as well, causing any initial displacement of the
center of mass to be exponentially amplified while the perturbation
works. Opposite to our models for the widths, this is an exact
prediction based solely on the GPE, and it shows that the parametrical
perturbation may also have a disturbing effect in the experiments. On
the other hand, a measure of this effect can give us information about
the intensity of dissipation and collision effects.

All of the preceding statements are based solely on the GPE. In a few
words, they include the existence of resonance regions both for the
widths and the center of mass, the shape and the location of those
regions, and its intensity as possible measure of damping. The failure of
any prediction should be interpreted as a failure of the GPE to describe
the condensate. Thus we provide with {\em simple} experimental checks
to perform a quantitative study of the regimes for which the GPE
properly describes the Bose-Einstein condensates in time dependent
traps.

Finally, we must mention that throughout this work we have concentrated
on regular motion regions in the parameter space. These regions can be
``safely'' reached in the experiments. There are many other cases where
chaos appears in the variational equations and complex behavior is seen
in the numerical simulations of Eq. (\ref{pura}). Although the study of
those disordered regions could be interesting from the nonlinear science
point of view they seem not to be of interest for Bose--Einstein
condensation since the exponential separation of nearby orbits which is
characteristic of chaotic behavior has been shown to induce
instabilities and take the system out of the regime where it can be
described using the mean field GP equation \cite{Castin2,Castin3}.

\acknowledgements

This work has been supported in part by the Spanish Ministry of
Education and Culture under grants PB95-0389, PB96-0534 and
AP97-08930807.

\end{document}